\documentclass[useAMS,usenatbib]{mn2e}
\usepackage{amssymb}
\usepackage{graphicx}
\usepackage{mathrsfs}
\usepackage{bm}
\newcommand\chandra{{\it Chandra}}

\newcommand\xmm{{\it XMM-Newton}}

\newcommand\s{{\rm~s}}
\newcommand\ks{{\rm~ks}}

\newcommand\kev{{\rm~keV}}
\newcommand\ev{{\rm~eV}}
\newcommand\kms{\ifmmode {\rm~km\ s}^{-1} \else ~km s$^{-1}$\fi}
\newcommand\Hunit{\ifmmode {\rm~km\ s}^{-1}\ {\rm Mpc}^{-1}
        \else ~km s$^{-1}$ Mpc$^{-1}$\fi}
\newcommand\ctssec{\ifmmode {\rm~count\ s}^{-1} \else ~count s$^{-1}$\fi}
\newcommand\ergsec{\ifmmode {\rm~erg\ s}^{-1} \else
        ~erg s$^{-1}$\fi}
 \newcommand\funit{\ifmmode {\rm~erg\ s}^{-1}\;{\rm cm}^{-2} \else
        ~ergs s$^{-1}$ cm$^{-2}$\fi}
\newcommand\phflux{\ifmmode {\rm~photon\ s}^{-1}\;{\rm cm}^{-2}
        \else   ~photon s$^{-1}$ cm$^{-2}$\fi}
\newcommand\efluxA{\ifmmode {\rm~erg\ s}^{-1}\;{\rm cm}^{-2}\;{\rm
        \AA}^{-1} \else ~erg s$^{-1}$ cm$^{-2}$ \AA$^{-1}$\fi}
\newcommand\efluxHz{\ifmmode {\rm~erg\ s}^{-1}\;{\rm cm}^{-2}\;{\rm
        Hz}^{-1} \else ~erg s$^{-1}$ cm$^{-2}$ Hz$^{-1}$\fi}
\newcommand\cc{\ifmmode {\rm~cm}^{-3} \else cm$^{-3}$\fi}
\newcommand\FWHM{\ifmmode {\rm~FWHM} \else ${\rm~FWHM}$\fi}
\newcommand\Msun{\ifmmode M_{\odot} \else $M_{\odot}$\fi}
\newcommand\Lsun{\ifmmode L_{\odot} \else $L_{\odot}$\fi}
\newcommand\ltsim{\raisebox{-.5ex}{$\;\stackrel{<}{\sim}\;$}}
\newcommand\gtsim{\raisebox{-.5ex}{$\;\stackrel{>}{\sim}\;$}}
\newcommand\hbeta{\ifmmode {\rm H}\beta \else H$\beta$\fi}
\newcommand\Kalpha{\ifmmode {\rm K}\alpha \else K$\alpha$\fi}
\newcommand\nh{\ifmmode N_{\rm H} \else N$_{\rm H}$\fi}

\title[State transition of the ULX NGC1313 X-1]{Associated Spectral and Temporal State Transition of the bright
  ULX NGC~1313~X-1} 
\author[G. C. Dewangan et al.]{G. C. Dewangan$^{1}$\thanks{gulabd@iucaa.ernet.in}, R. Misra$^{1}$\thanks{rmisra@iucaa.ernet.in}, A. R. Rao$^{2}$\thanks{arrao@tifr.res.in} and
  R. E. Griffiths$^{3}$\thanks{griffith@seren.phys.cmu.edu} \\
$^{1}$IUCAA, Post Bag 4, Ganeshkhind, Pune 411 007, India \\
$^{2}$Department of Astronomy \& Astrophysics, Tata
  Institute of Fundamental Research, Homi Bhabha Road, Mumbai, 400005
  India \\
$^{3}$Department of Physics, Carnegie Mellon University, 5000 Forbes Avenue,
  Pittsburgh, PA 15213 USA}
\begin{document}

\date{Submitted 2009 May}

\pagerange{\pageref{firstpage}--\pageref{lastpage}} \pubyear{????}
\maketitle
\label{firstpage}

\begin{abstract}
  Stellar mass black hole X-ray binaries exhibit X-ray spectral states
  which also have distinct and characteristic temporal properties.
  These states are believed to correspond to different accretion disk
  geometries.  We present analysis of two long \xmm{} observations of
  the Ultra-Luminous X-ray source (ULX) NGC~1313 X-1, which reveal
  that the system was in two different spectral states.  While 
  spectral variations have been observed in this source before, this
  data provides clear evidence that the spectral states also have
  distinct temporal properties.  With a count rate of $\sim 1.5$
  counts/s and a fractional variability amplitude of $\sim 15\%$, the
  ULX was in a high flux and strongly variable state in March 2006.
  In October 2006, the count rate of the ULX had reduced by a factor
  of $\sim 2$ and the spectral shape was distinctly different with the
  presence of a soft component. No strong variability was detected during
  this low flux state with an upper limit on the amplitude $< 3\%$.
  Moreover, the spectral properties of the two states implies that the
  accretion disk geometry was different for them. The low flux state
  is consistent with a model where a standard accretion disk is
  truncated at a radius of $\sim 17$ Schwarzschild radius around a
  $\sim 200 M_\odot$ black hole. The inner hot region Comptonizes
  photons from the outer disk to give the primary spectral
  component. The spectrum of the high flux state is not compatible
  with such a geometry. Instead, it is consistent with a model where a
  hot corona covers a cold accretion disk and Comptonizes the disk
  photons. The variability as a function of energy is also shown to be
  consistent with the corona model.  Despite these broad analogies
  with Galactic black hole systems, the spectral nature of the ULX is
  distinct in having a colder Comptonizing temperature ($\sim 2$ keV)
  and higher optical depth ($\sim 15$) than what is observed for the
  Galactic ones.
\end{abstract}


\section{Introduction}
Ultra-luminous X-ray sources (ULXs) are off-nuclear, compact X-ray
sources with luminosities exceeding $\sim 10^{39}{\rm~erg~s^{-1}}$.
The nature of ULXs continues to be an enigma, since their isotropic
high energy output surpasses the Eddington limit of even the most
massive stellar mass black holes, sometimes by large factors.  One
individual ULX may outshine the rest of the galaxy in the high energy
X-ray band.  Several models have been proposed to explain the high
luminosities of ULXs. They may be powered by ``intermediate mass black
hole (IMBH)'' with masses $M_{BH} \simeq 10^2 - 10^4{\rm~M_{\odot}}$
\citep[e.g.,][]{2005tmgm.meet..530C} which bridge the gap between
stellar mass black holes in X-ray binaries (XRBs) and super-massive
black holes in active galactic nuclei.  On the other hand, they may be
XRBs with anisotropic emission \citep{2001ApJ...552L.109K} or beamed
XRBs with relativistic jets directly pointing towards us, i.e., scaled
down versions of blazars \citep{1999ARA&A..37..409M}. Alternatively,
they may be stellar mass black holes with super-Eddington accretion
rates \citep{2002ApJ...568L..97B, 2008MNRAS.385L.113K}.

In the absence of a direct measurement of the black hole mass in ULX,
indirect evidence is provided by the spectral and temporal
properties of these sources.  Qualitative and quantitative
similarities (and differences) between them and Galactic X-ray
binaries harboring stellar mass black holes can provide insight into
their nature. The discovery of orbital modulations from the bright ULX
M~82 X-1 \citep{2006ApJ...646..174K,2007ApJ...669..106K} and possibly
from a ULX in IC~342 \citep{2001ApJ...561L..73S} implies that they too
are binary systems.

Galactic X-ray binaries exhibit spectral states \citep[e.g. see][for a
review] {2004PThPS.155...99Z}.  Detailed spectral modeling of these
systems reveal that the accretion disks have different geometries for
different states. In the high soft state, the spectra are dominated by
emission from a cold accretion disk which extends probably to the
inner most stable orbit. A corona or active region above the cold
disk comptonizes disk photons to produce hard X-rays.  In the low hard
state, the disk is truncated at a larger radius and the X-rays
originate from an hot inner disk. In the very high state, the geometry
is probably similar to that of the high state, but here the flux from
the corona is nearly equal to that of the disk. X-ray binaries are
variable on a wide range of time-scales and exhibit broad band noise
as well as narrow features called Quasi-periodic oscillations (QPO).
Spectral state transitions are always associated with distinct changes
in the rapid temporal behavior of the systems.

Both \chandra{} and \xmm{} observations of ULXs have sometimes shown
soft X-ray excess emission which has been interpreted as the optically
thick emission from thin accretion disks with temperatures in the
range of $\sim 100-300\ev$, which indicates that they are IMBH
accreting at sub-Eddington ($\sim 0.1L_{Edd}$) rates
\citep[e.g.,][]{2003ApJ...585L..37M,2003Sci...299..365K,2004ApJ...607..931M}.
While this suggests that ULX may be scaled up versions of X-ray
binaries, it should be noted that there are important spectral
differences. Unlike X-ray binaries, a high energy spectral curvature
has been detected in the the high quality X-ray spectra of a number of
bright ULXs
\citep{2006ApJ...638L..83A,2006MNRAS.368..397S,2006MNRAS.365..191G,2006ApJ...641L.125D,2007Ap&SS.311..203R}. This
spectral curvature is consistent with strong, nearly saturated
Comptonization. Such strong Comptonization models have been invoked to
explain the very high state of black hole binaries such as
XTE~1550-564 \citep{2006MNRAS.371.1216D}. However, the temperature
derived for the electron plasma ($\sim 2\kev$) is significantly
smaller than that of X-ray binaries ($> 20$ keV).

Another striking similarity between ULX and X-ray binaries is the
presence of significant variability power as well as features in the
power density spectra such as QPOs and breaks
\citep{2003ApJ...586L..61S,2006ApJ...641L.125D,2006ApJ...637L..21D,2006MNRAS.365.1123M,2007ApJ...660..580S}. The
QPO and break frequencies are found to be smaller than those of X-ray
binaries which is taken as evidence that ULX harbor IMBH. For some
sources, like the bright ULX Holmberg II X-1, the variability is
absent or very weak ($< 2\%$) \citep{2006MNRAS.365..191G}.

Studies of a collection bright ULX in nearby galaxies reveal that for
some ULX a power-law spectral fit (similar to the low/hard state of
X-ray binaries) is adequate while others require a thermal component
(similar  to the high/soft state).  The temperature of the thermal
component is either $\sim 0.2$ or $\sim 2\kev$ suggestive of a further
subdivision \citep{2006ApJ...649..730W,2007ApJ...664..458D}. These
studies do not indicate whether these spectral differences represent
different spectral states or if ULX themselves are a heterogeneous
class comprising different kinds of sources.  Long term spectral
variability, which is interpreted as state transitions similar to
X-ray binaries, has been reported for several ULX.  Repeated
\chandra{} observations of the Antennae have shown hardness ratio
changes in several ULXs. Transition from a high soft state to a low
hard state has been observed from two ULXs in IC~342
\citep{2001ApJ...547L.119K} and Holmberg IX X-1
\citep{2001ApJ...556...47L}.  High/hard to low/soft spectral
variability has also been observed from Holmberg II X-1
\citep{2004ApJ...608L..57D}.

However, these spectral variations have not been shown to be
associated with temporal behavior changes and hence their
identification as spectral state transitions similar to those observed
in X-ray binaries remains uncertain. Moreover, the spectral variations
recorded do not necessarily imply a change in the accretion disk
geometry.  Well constrained spectral parameters are needed for such an
inference. Hence, good high quality data from two long observations of
a source undergoing such changes are needed to ascertain whether ULX
undergo true state transitions.

NGC~1313 is a spiral galaxy and is located at a distance of
$4.13\pm0.11{\rm~Mpc}$ \citep{2002AJ....124..213M}. The galaxy hosts
three bright and well separated X-ray sources: two ULXs X-1 and X-2,
and a supernova remnant SN~1978K. NGC~1313 X-1 is one of the first ULX
observed by \xmm{} to show soft X-ray excess emission that led to the
$\sim 1000M_{\odot}$ IMBH interpretation
\citep{2003ApJ...585L..37M}. The ULX belongs to the low-temperature,
high-luminosity class in the luminosity versus temperature ($L-kT$)
diagram of bright ULXs in nearby galaxies \citep{2005ApJ...633.1052F}.
\cite{2006ApJ...650L..75F,2007ApJ...660L.113F} have utilized the first
14 \xmm{} observations and reported long term spectral variations of
the two ULXs NGC~1313 X-1 and X-2. For X-1, they fitted the spectra
with an absorbed power-law model and reported a correlation between
the photon index, $\Gamma$ and luminosity. An additional soft black
body component was required for six of the observations. The short
exposures of these observations (in the range of $9-41\ks$), did not
allow for detailed spectral and temporal studies.

Here we present detailed temporal and spectral properties of the ULX
using two  \xmm{} observations of $123$ and $20\ks$ . We
describe the \xmm{} observations and the spectral and temporal
analysis in \S{2}. In \S{3}, the accretion disk geometries inferred
from the analysis are described.  We summarize and discuss the results
in \S{4}.

\begin{figure*}
  \centering
  \includegraphics[width=8.5cm]{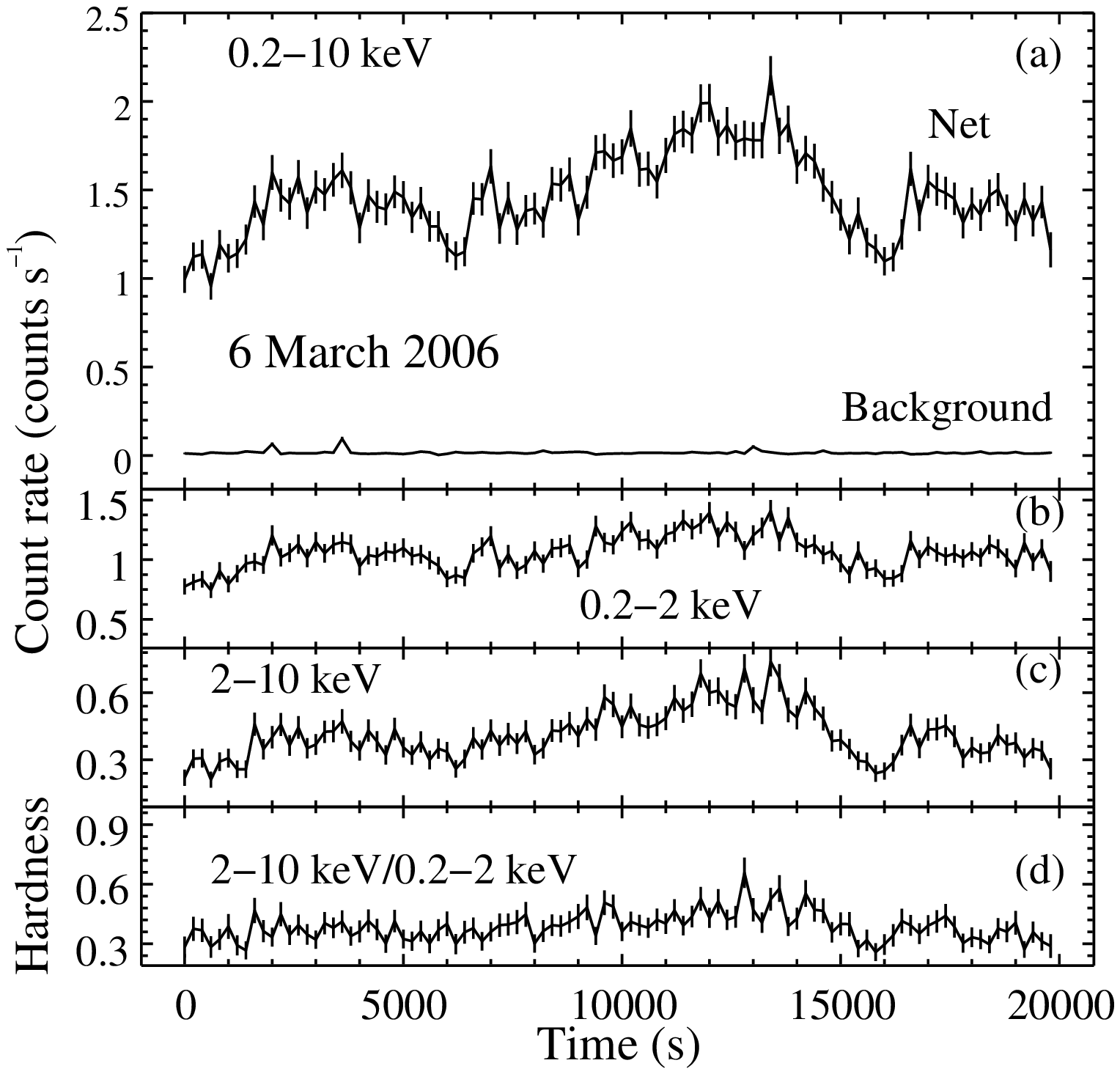}
  \includegraphics[width=8.5cm]{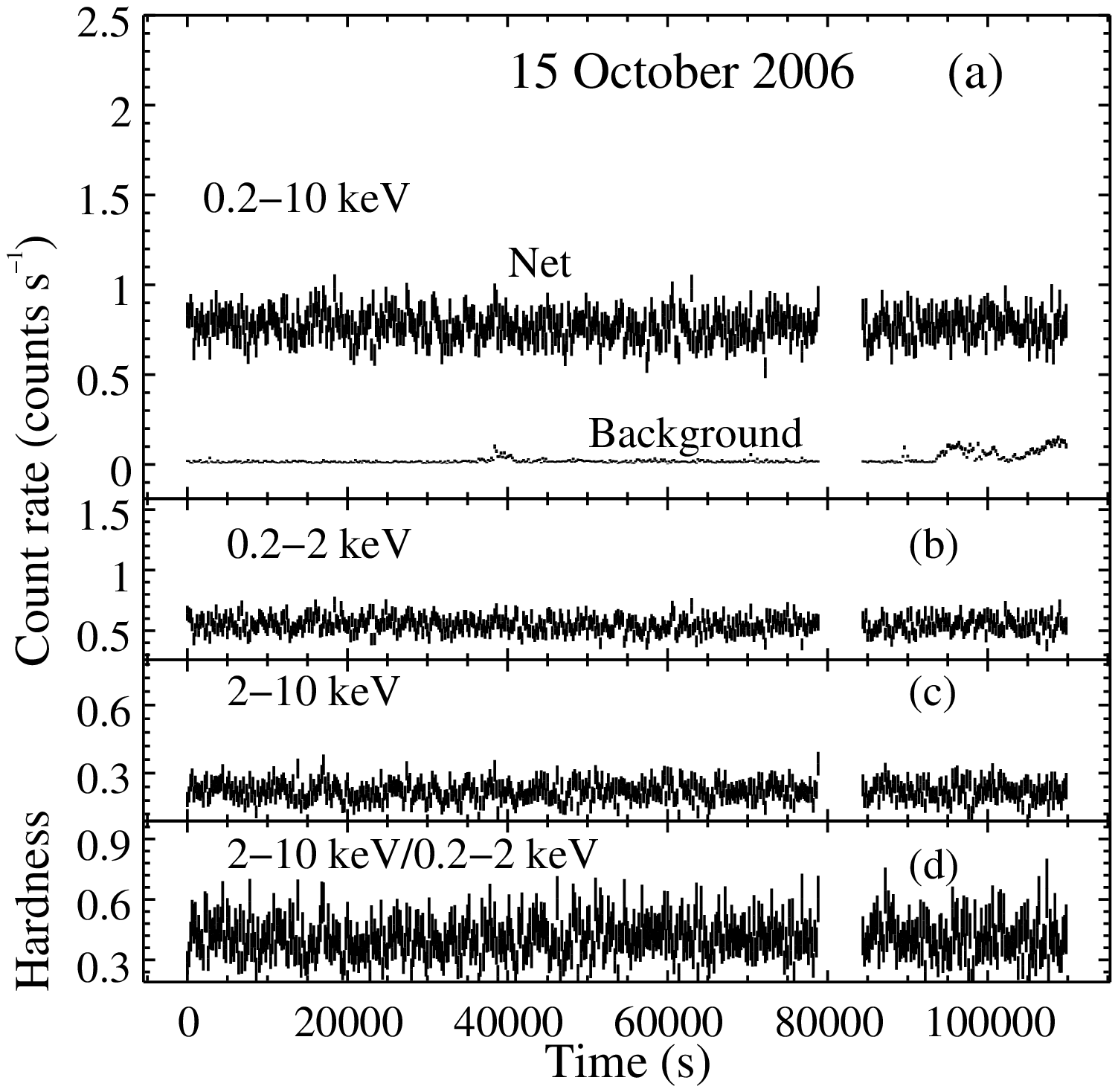}
  \caption{The background corrected EPIC-pn lightcurves of NGC~1313
    X-1 binned with $200\s$ in the (a) $0.2-10\kev$, (b) $0.2-2\kev$,
    (c) $2-10\kev$, and (d) the hardness ratio derived from the
    observations of 6 March 2006 ({\it left panel}) and 15 October
    2006 ({\it right panel}). The relative contribution of background
    emission in the source extraction region is also shown for the
    $0.2-10\kev$ band. }
  \label{f1}
\end{figure*}


%
\section{Observations and Data Reduction} \label{obs_red}

We observed NGC~1313 with \xmm{} for a $123$ ks long exposure on 15
October 2006.  NGC~1313 was also observed on 6 March 2006 with an
exposure of $21.8\ks$. We used SAS 8.0.1 with updated calibration to
process and filter both EPIC-pn and MOS data.  The long observation of
15 October 2006 was partly affected with low amplitude flaring
particle background towards the end of the exposure.  Cleaning of the
flaring particle background resulted in $89.3\ks$ data. The
observation of 6 March 2006 was not affected by particle background.
For temporal analysis we selected events with patterns $0-12$ and used
either continuous exposure free of particle background or corrected
for the background contribution.  For spectral analysis, we selected
the events with pattern $0-4$ (single and double) for the EPIC-pn and
$0-12$ for the MOS and excluded events adjacent to the bad pixels with
FLAG=0. For the observation of March 2006, the source extension
overlaps with the EPIC-pn chip gap and a large fraction of events have
uncertain pattern. These events with poor spectral calibration were
excluded from the spectral analysis but were included in the temporal
analysis.

\begin{table*}
    \caption{Variability
    properties of NGC~1313 X-1 \label{tab1}} 
\begin{tabular}{lcccccc} \hline
    Energy band & \multicolumn{3}{c}{6 March 2006} & \multicolumn{3}{c}{15 October 2006} \\
                &  Source & Background & $F_{var}$($\%$)$^{1}$ & Source & Background & $F_{var}$($\%$)$^{1}$ \\
    & ($\rm counts~s^{-1}$) & ($\rm counts~s^{-1}$) & & ($\rm
    counts~s^{-1}$) & ($\rm counts~s^{-1}$)  \\ \hline 
  $0.2-10\kev$ & $1.47\pm0.09$ & $0.016\pm0.004$ &$14.6\pm0.6$  & $0.77\pm0.07$ & $0.026\pm0.006$  & $2.3_{-1.9}^{+0.9}$  \\
  $0.2-2\kev$  & $1.06\pm0.08$ & $0.009\pm0.003$ & $11.0\pm0.8$ & $0.55\pm0.06$ & $0.010\pm0.004$  & $2.9_{-1.7}^{+1.0}$ \\
  $2-10\kev$   & $0.41\pm0.05$ & $0.006\pm0.003$ & $23.7\pm1.2$ & $0.22\pm0.04$ &  $0.013\pm0.004$   & $<4.3$ \\ \hline
  \end{tabular}

  $^{1}${Calculated from background subtracted lightcurve
    with $200\s$ bins.}
\end{table*}

\begin{figure*}
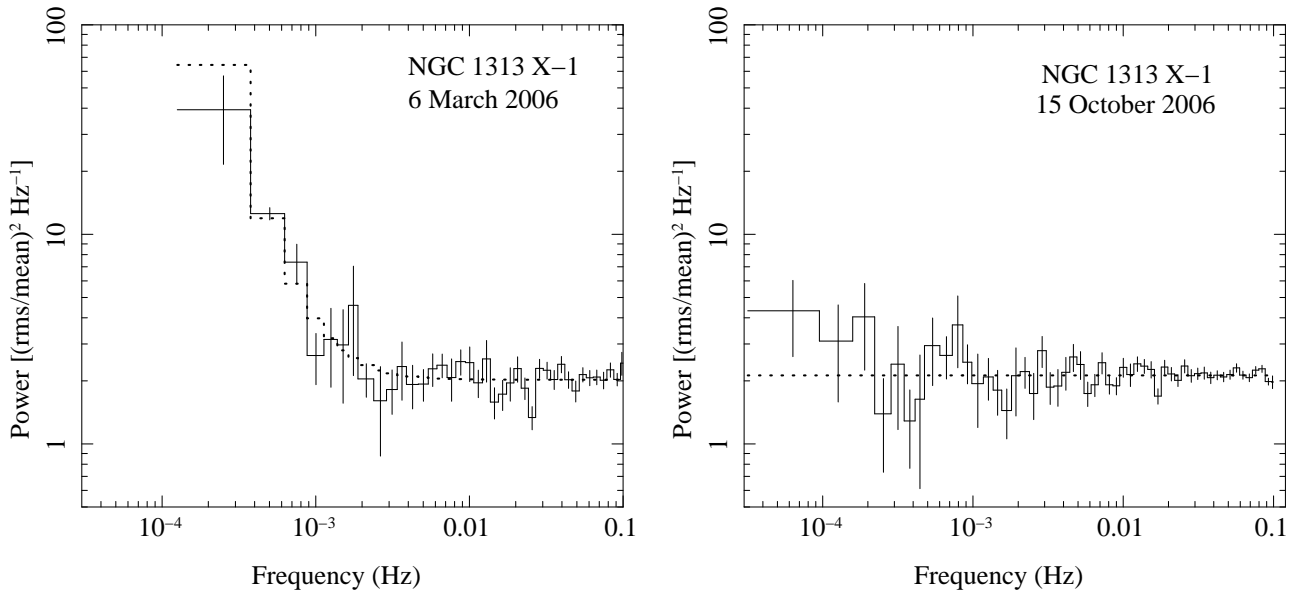

  \centering
  \includegraphics[width=8.0cm,angle=-90]{x1_psd_6mar06_bs5s_int5_m1p1_plaw_cons.ps}
  \includegraphics[width=8.0cm,angle=-90]{x1_psd_rbs5s_m1p1_int5_const.ps}
  \caption{Power density spectra of NGC~1313 X-1 derived from the
    EPIC-pn lightcurves obtained on 6 March 2006 and 15 October 2006}
  \label{f2}
\end{figure*}

\subsection{Temporal Analysis}
We extracted two sets of lightcurves from each of the observations
with bin sizes of $200\s$ and $0.5\s$. A circular region of
$40\arcsec$ centered at the source position was used to extract the
EPIC-pn source lightcurves. We also extracted background lightcurves
using four circular regions of radii $40\arcsec$ in the source-free
areas and close to the source and used them to correct the source
lightcurves for background contribution. In Figure~\ref{f1}, we show
the background corrected lightcurves of NGC~1313 X-1 in the
$0.2-10\kev$, $0.2-2\kev$ and $2-10\kev$ bands. For comparison, we
have also plotted the background contribution in the source extraction
region. Evidently background contribution is nearly constant and
negligible throughout the observations.

As can be seen in Fig.~\ref{f1} ({\it left panel}), NGC~1313 X-1
showed strong X-ray variability throughout the observation of 6 March
2006. The most remarkable variability in the $0.2-10\kev$ band
occurred at an elapsed time of $\sim13000\s$ when the net count rate
decreased by a factor of about two in $3000\s$.  The hardness ratio,
calculated as the ratio of count rates in the $0.2-2\kev$ and
$2-10\kev$ band is also variable and correlated with the
lightcurve. Thus, at high flux, the ULX has a harder spectrum.

In contrast to the $20\ks$ observation on 6 March 2006, the long
observation of 15 October revealed that  ($i$) the ULX was in a low
flux state, X-ray emission dropped by a factor of about $\sim 2$ , and
($ii$) the emission was remarkably steady throughout the $\sim 120\ks$
exposure. The soft and hard band emission and the hardness ratio also
remained steady, implying that the spectrum of the ULX did not change
substantially (see Fig.~\ref{f1}, {\it right panel}).  Henceforth,
throughout this work we refer to the 6 March observation as the high
state and the 15 October one as the low state. 

To quantify the variability, we have calculated the intrinsic source
variability expressed in terms of the fractional variability amplitude
($F_{var} = ({\sigma_{XS}^2}/{\overline{x}^2})^{1/2}$, where
$\sigma_{XS}^2$ is the variance after subtracting the contribution
expected from measurement errors, and $\overline{x}$ is the mean count
rate \citep[see][and references therein]{2003MNRAS.345.1271V}. We used
the background corrected lightcurves binned with $200\s$ to calculate
$F_{var}$, listed in Table~\ref{tab1}. While the measured $F_{var}$ is
consistent with no intrinsic variability during the low state,
significant intrinsic variability is seen during the high state.  We
have also derived the power density spectra using the EPIC-pn
lightcurves binned with $5\s$. For the low state, we used the first
$\sim79\ks$ continuous exposure in order to avoid the data gap and the
flaring particle background seen towards the end of the observation
(see Fig.~\ref{f1}). The light curves were divided into five equal
segments and the PSDs of each segment was computed.  The five
individual PSDs were averaged and the errors were assigned to be the
standard deviation at each frequency bin. The resultant PSDs (after
binning) are shown in Figure~\ref{f2}. The PSD shows red-noise
behavior at low frequencies for the hard state, while it is flat
during the low one.  A constant model fitted to the PSD for the low
state resulted in a minimum $\chi^2 = 59.3$ for $52$ degrees of
freedom (dof). The best-fit constant power density is $2.12\pm0.04$
which is slightly higher than the power density of $2$ expected from
pure Poisson noise arising from photon counting statistics. The
marginal higher power density could arise from slight variation
in the background level (see Fig.~\ref{f1} {\it right panel}). Thus,
we measure no significant intrinsic variability, consistent with the
measured $F_{var}$ (see Table~\ref{tab1}). As expected, a constant
power density model fit to the PSD of the high state resulted in a
statistically unacceptable fit ($\chi^2/dof = 211.5/38$). Addition of
a power-law component ($\propto \nu^{-\alpha}$) improved the fit
significantly ($\chi^2/dof = 47.7/36$) with $\alpha = -2.25\pm0.30$.
Interestingly a broken power-law model further improved the fit
marginally ($\chi^2/dof = 40.9/34$). However, the parameters of the
model are not well constrained and hence we conclude that the data is
suggestive of a break, but no concrete inference can be made.

\subsection{Spectral Analysis}
We extracted EPIC-pn and MOS spectra from the two observations using
circular regions with radii of $40\arcsec$ centered at the position of
NGC~1313 X-1. We also extracted background spectra using appropriate
nearby circular regions free of sources. We created spectral response
files using the SAS tasks {\tt rmfgen} and {\tt arfgen}.  Spectral
bins were chosen such that there was a minimum of $20$ counts per
spectral channel for the pn and the MOS data of March 2006 (high
state). The high signal-to-noise EPIC-pn, MOS1 and MOS2 data extracted
from the deep observation of October 2006 (low state) were grouped to
minimum counts of $200$, $50$ and $50$, respectively.  These spectra
were analyzed with {\tt ISIS 1.4.9}. All spectra were fitted in the
energy range $0.3-10\kev$ and the errors on the best-fit spectral
parameters are quoted at a $90\%$ confidence level.

\begin{figure}
\centering
  \includegraphics[width=7cm,angle=-90]{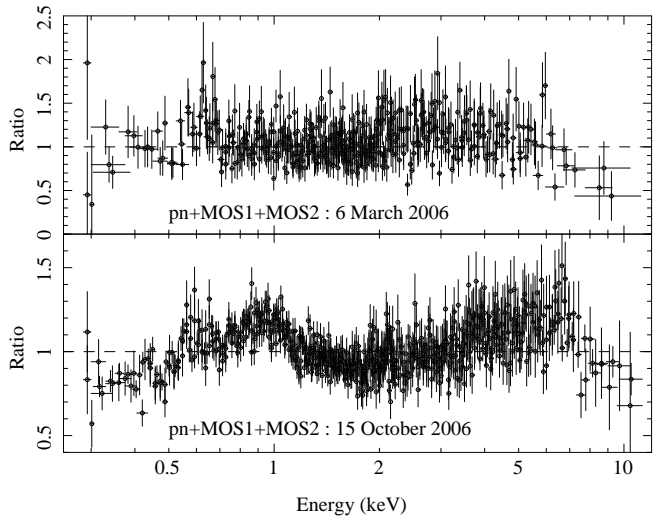}
  \caption{Ratios of observed pn+MOS data obtained on 6 March 2006
    ({\it top panel}) and 15 October 2006 ({\it bottom panel}) and the
    best-fitting absorbed power-law models.}
  \label{f3}
\end{figure}

\begin{figure}
\centering
  \begin{center}
    \includegraphics[width=8.0cm,angle=-90]{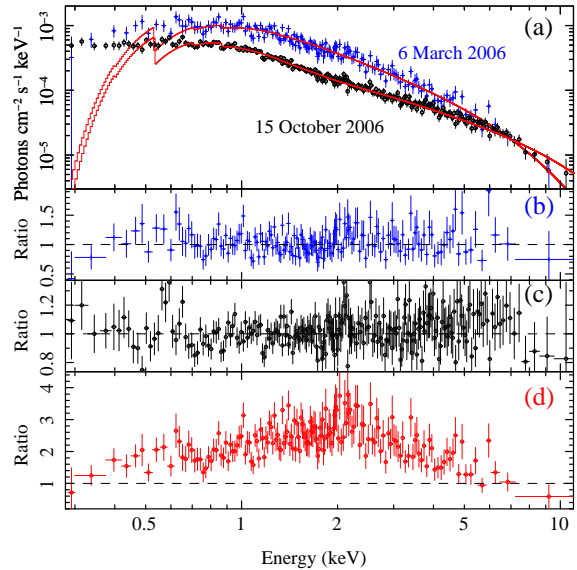}
  \end{center}
  \caption{Results of spectral fitting to the EPIC-pn data obtained on
    high (6 March 2006) and low (15 October 2006) states. (a) {\it
      Lower plot:} Observed data for the high state and the best-fit
    model - thermal Comptonization in a cool, optically thick medium
    Unfolded (nthcomp); {\it Upper plot:} Observed data for the low
    state and the best-fit model MCD+nthcomp.  (b) Deviations of the
    observed data of the high state from the best-fit absorbed nthcomp
    model. (c) Deviations of the observed data for the low state from
    the best-fit absorbed MCD+nthcomp model. (d) Difference in the
    shape of the spectra for the high and low states. The plot shows
    the deviations of the data of the low state from the best-fit
    model to the high state data but renormalized to match the fluxes.
  }
  \label{f4}
\end{figure}

We begin with the spectral analysis of the low state. A simple
absorbed power law (PL) model was fitted individually to the pn and
MOS spectra.  Examination of the individual fits showed the photon
indices and absorption columns obtained for three data sets were
similar within errors and the fits had similar residuals. Therefore,
we fitted the pn and MOS data jointly and used an overall
normalization constant to account for possible differences in source
extraction areas or calibration uncertainties. The simple power-law
model resulted in an statistically unacceptable fit ($\chi^2/dof =
1484.0/674$). The ratio of the pn+MOS data and the power-law model is
shown in Fig.~\ref{f3}({\it bottom panel}). This plot clearly shows a
broad hump or a cutoff near $6\kev$ and a soft X-ray hump or excess
emission in the $0.6-1.2\kev$ band. Similar spectra have been reported
earlier for several ULXs
\citep{2005ApJ...633.1052F,2006MNRAS.368..397S,2006ApJ...641L.125D}.
Addition of a multicolor disk blackbody (MCD) component to the PL
model improved the fit significantly ($\chi^2/dof =
807/672$). Replacing the PL with a cutoff PL further improved the fit
($\chi^2/dof = 717.3/671$).  This kind of a high energy cut-off is
common among bright ULXs and is well described by thermal
Comptonization in a relatively cool and optically thick
plasma. \citep{2006ApJ...638L..83A,2006MNRAS.368..397S,2006MNRAS.365..191G,2006ApJ...641L.125D,2007Ap&SS.311..203R}.
To test such a scenario, we replaced the cutoff PL model with the
thermal Comptonization model (nthcomp) described by
\citet{1996MNRAS.283..193Z} and \citet{1999MNRAS.309..561Z}.  The free
parameters of the nthcomp model are the asymptotic power-law index
($\Gamma_{thcomp}$), electron temperature ($kT_e$) and the seed photon
temperature ($kT_{S}$). The electron scattering optical depth ($\tau$)
can be calculated from the asymptotic power-law photon index
($\Gamma_{compth}$) and electron temperature ($kT_e$) as follows
\begin{equation}
  \alpha = \left[\frac{9}{4} + \frac{1}{(kT_e / m_e c^2) \tau (1 + \tau / 3)}\right]^{1/2} - \frac{3}{2}
\end{equation}
with $\Gamma_{thcomp} = \alpha + 1$ \citep{1980A&A....86..121S}.
Since a soft X-ray excess emission, described by an MCD, is clearly
detected, it is simplistic to assume that this component provides the
seed photons for the Comptonization. Therefore, the temperature of the
soft excess and the seed photons were kept the same and varied
together. The shape of the soft component was assumed to be that of a
MCD.  The MCD+nthcomp model provided a good fit ($\chi^2/dof =
717.9/671$) and the best-fit parameters are reported in Table
(\ref{tab2}). To confirm that the results do not depend on the
Comptonization model used, we have fitted the data using the XSPEC
model compTT
\citep{1994ApJ...434..570T,1995ApJ...449..188H,1995ApJ...450..876T}
instead of nthcomp and obtained a $\chi^2/dof = 716.3/671$ with
similar parameter values. In Fig.~\ref{f4}, we show the unfolded
EPIC-pn spectrum, the best-fitting absorbed MCD+ncompth model in panel
(a) ({\it upper curve}), and the deviations of the data from the model
in panel (c). For clarity we show only the EPIC-pn data. Unlike XPSEC,
the unfolded spectrum in ISIS is derived in a model-independent way as
follows:
\begin{equation}
  f_{unfold} (I) = \frac{[C(I) - B(I)]/\Delta t}{\int {R(I,E)A(E)dE}},
\end{equation}
where C(I) is the number of total counts in the energy bin $I$, $B(I)$
is the number of background counts, $\Delta t$ is exposure time,
$R(I,E)$ is the normalized response matrix and $A(E)$ is the effective
area at energy $E$.  This definition produces a spectrum that is
independent of the fitted model. The best-fit model has been
over-plotted at the internal resolution of the ancillary response
function. The spectrum and the model are expected to match only in the
particular case of unit response matrix
\citep{2005Ap&SS.300..159N}. Hence there is deviation between the
model and the data points at low energies. Note that the {\it model
  independent} unfolded data points clearly shows a curvature at high
energies, which signifies that the curvature is real and not a
modeling artifact.

\begin{table}
\begin{minipage}{140mm}
   \caption{Best-fit spectral
    model parameters for NGC~1313 X-1.
    \label{tab2}}
\begin{tabular}{lllll}
\hline 
Parameters & 6 March 2006  & 15 October 2006 \\
           & nThComp & MCD+nThComp \\ \hline
  $N_H$($10^{21}{\rm~cm^{-2}}$) & $2.50$(fixed)    & $2.50_{-0.15}^{+0.16}$ \\
  & & \\
  $kT_{in}$($\ev$)$^{a}$          & --                 & $227$ \\
  & & \\
  $f_{MCD}^{b}$     & --  & $1.24_{-0.12}^{+0.15}\times10^{-12}$          \\
  & & \\
  $kT_{S}$($\ev$)$^{c}$ & $122\pm15$ & $227_{-12}^{+14}$ \\
  & & \\
  $kT_{e}$($\kev$)$^{d}$  &$1.67_{-0.19}^{+0.28}$ & $2.27_{-0.15}^{+0.18}$\\
  & & \\
  $\Gamma ^{e}$                & $2.02_{-0.04}^{+0.05}$ & $1.67\pm0.03$ \\
  $f_{comp}^{f}$     & $7.51_{-0.23}^{+0.24}\times10^{-12}$  &  $(3.1\pm0.1)\times10^{-12}$  \\
  & & \\
  $f_X^{g}$         & $7.51_{-0.23}^{+0.24}\times10^{-12}$  & $4.34_{-0.18}^{+0.21}\times10^{-12}$ \\
  & & \\
  $L_{bol}^{h}$  & $17.9\times10^{39}$   & $11.9\times10^{39}$  \\
  & & \\
  $\chi^2_{min}$/dof         & $421.4/439$          & $717.9/671$ \\ \hline
\end{tabular}
\end{minipage}
  $^{a}${Inner disk temperature of disk blackbody model
    which is tied to $kT_S$.} \\
    $^{b}${$0.3-10\kev$ flux of the MCD component in ${\rm
      ergs~cm^{-2}~s^{-1}}$.}  \\
$^{c}${Temperature of the
    seed photon source of the Comptonization model } \\
    $^{d}${Hot electron plasma temperature.} \\
  $^{e}${Photon Index of the Comptonization model.} \\
  $^{f}${$0.3-10\kev$ flux of the Comptonization model in
    ${\rm ergs~cm^{-2}~s^{-1}}$.} \\
  $^{g}${Unabsorbed $0.3-10\kev$ flux in ${\rm
      ergs~cm^{-2}~s^{-1}}$.} \\
  $^{h}$Unabsorbed bolometric luminosity in ${\rm
      ergs~s^{-1}}$. 

\end{table}

We then proceed to the spectral analysis of the pn and MOS data of the
high state.  As mentioned earlier in section \ref{obs_red}, the
extension of the ULX in the EPIC-pn image overlaps with the chip gap,
hence a good fraction of the events at high energies were discarded as
these events have poorly determined patterns. Consequently, the
EPIC-pn and the MOS data have comparable signal-to-noise.  Simple
absorbed power-law model fitted separately to the pn, MOS1 and MOS2
data resulted in similar photon indices and we did not find
significant differences in the residuals, therefore we present results
based on joint spectral fitting of the pn and MOS data as before. The
simple power law model resulted in a minimum $\chi^2 = 455.3$ for
$440$ dof.  With a reduced $\chi^2 = 1.03$, a simple power-law model
seems adequate and does not seem to warrant exploration of more
complex models. Indeed \cite{2006ApJ...650L..75F} were able to model
several similar quality data of this source with a simple
power-law. However, the analysis of better quality data described
above suggests that the spectrum of the source could be more
complex. Moreover, as we show below, there is evidence for a more
complex spectral shape even for this statistically poorer data set. We
have plotted the ratio of the pn+MOS1+MOS2 data and the best-fit PL
model in Figure~\ref{f3}({\it upper panel}). The absorbed power-law
model is a good description of the data below $\sim 6\kev$. There is a
likely cutoff or curvature at high energies. To verify the possible
high energy turnover of the X-ray spectra, we replaced the PL
component by a cut-off PL. The absorbed cut-off PL model yielded
$\Delta \chi^2 = -19.9$ for one additional parameter as compared to
the absorbed PL model.  This is an improvement at a significance level
of $>99.99\%$ based on the Maximum Likelihood Ratio test.  The
absorbed nthcomp model provided a similar fit with $\chi^2/dof =
421.4/439$. In Figure~\ref{f4}, we show the unfolded EPIC-pn spectrum,
the best-fitting absorbed ncompth model in panel (a) marked as 6 March
2006 and the deviations in panel (b).  Note that, similar to the low state
data set, the {\it model independent} unfolded data points clearly
show a curvature at high energies which again signifies that the
curvature is real and not a modeling artifact. It is also clear from
the model independent unfolded data that both the flux and spectral
shape varied between the two data sets. This is highlighted in panel
(d) of the Figure which shows the deviations of the low state data
from the renormalized best fit model of the high state. We have also
investigated the possible presence of a disk component similar to that
observed in the low state. Addition of an MCD component to the nthcomp
model with MCD $kT_{in}$ tied to the nthcomp $kT_S$ did not improve
the fit. We calculated an upper-limit of $L_{MCD} <
1.4\times10^{39}{\rm~ergs~s^{-1}}$ for the bolometric luminosity of
the MCD component in the $0.001-100\kev$ band. We also fixed the MCD
$kT_{in}$ at the value obtained for the low state and calculated the
upper limit, $L_{MCD} < 6.3\times10^{39}{\rm~ergs~s^{-1}}$.

\section{Accretion disk Geometry}

As revealed in Table \ref{tab2}, the low state spectrum of the source
can be described with a thermal Comptonization and a disk black body
model.  The luminosity of the disk black body (i.e. the MCD model) is
$L_{MCD} \sim 4 \times 10^{39}$ ergs/s, while that of the thermal
Comptonization is $\sim 7.7 \times 10^{39}$ ergs/s. Using the XSPEC
model function for the model ``nthcomp'',one can estimate the
amplification factor, $A \equiv L_c/L_s \sim 4.8$, where $L_c$ and
$L_s$ are the luminosities of the Comptonization component and the
seed photon input. This allows an estimate of $L_s \sim 1.6 \times
10^{39}$ ergs/s. This is consistent with an accretion disk geometry
where a standard cold accretion disk is terminated at radius $R_{tr}$
and the inner region is an hot plasma. The inner hot plasma
Comptonizes photons from the outer disk producing high energy photons.
For consistency, the seed photon spectral shape for the Comptonization
model is taken to be a disk black body emission at the same
temperature as the disk emission observed.  In this model, the
fraction of the cold disk photons that enter the inner hot region is
related to the solid angle which the hot region makes with the outer
disk. The observed fraction of the luminosity of the seed photon
component as compared to the disk luminosity, $L_{S}/L_{MCD} \sim 0.4
$ suggests a solid angle of $\Delta \Omega \sim 0.4 \times 2 \pi \sim
0.8 \pi$. Note that $L_{S}/L_{MCD} \ltsim 1$ is a consistency check
for the geometry proposed. In this geometry, it would be difficult to
reconcile $L_S > L_{MCD}$ and $L_S << L_{MCD}$ would require an
unphysical small subtending solid angle.

The normalization of the disk component provides an estimate of the
transition radius
\begin{equation}
  R_{tr} ~ \sim ~ 10^9 ~ {\rm cm} ~~ \left(\frac{\kappa}{1.7}\right)^2 ~ \left(\frac{{\rm {cos}}i}{0.5}\right)^{-1/2} 
\end{equation}
where $i$ is the inclination angle and $\kappa$ is the color factor.
In the standard accretion disk theory \citep{1973A&A....24..337S}, the
luminosity of a truncated disk is given by
\begin{equation}
  L_{MCD} ~ \sim ~\frac{3}{2} \frac{G M \dot M}{R_{tr}}
\end{equation}
where $M$ is the mass of the black hole and $\dot M$ is the accretion
rate.  The total bolometric luminosity, $L_T = (L_C - L_S) + L_{MCD}$
can be expressed as $L_T = \eta \dot M c^2$ where $\eta$ is the
efficiency of the system. Thus, the mass of the black hole can be
estimated to be
\begin{eqnarray}
  M ~ & \sim ~ & \frac{2c^2}{3G} ~ \eta ~ R_{tr} ~ \frac{L_{MCD}}{L_T} \nonumber \\
  & \sim ~ & 200 ~ M_\odot ~ \left(\frac{\eta}{0.1}\right) ~ \left(\frac{R_{tr}}{10^9 {\rm cm}}\right) ~ \left(\frac{L_{MCD}/L_T}{0.43}\right)
  \label{bhmass} 
\end{eqnarray}

The Eddington luminosity for a $200 M_\odot$ solar mass black hole is
$L_{Edd} = 2.5 \times 10^{40}$ ergs/s and hence the source is
radiating at an Eddington ratio of $L_T/L_{Edd} \sim 0.37$. The
Schwarzschild radius turns out to be $ r_s \equiv 2GM/c^2 \sim 6
\times 10^7$ cm, making the transition radius $R_{tr} \sim 17 r_s$.

Thus the spectral shape of the low state is totally consistent with a
scenario where a $\sim 200 M_\odot$ solar mass black hole is
surrounded by a standard accretion disk which is truncated at $\sim 17
r_s$. The disk produces a multi temperature emission which is observed
as a soft component. The inner hot region is geometrically thick and
subtends a solid of angle of $\sim 0.8 \pi$ to the disk outside. This
allows it to Comptonize photons from the outer disk producing the main
Comptonized component.  The system is radiating at a Eddington ratio
$\sim 0.4$ with a standard radiation efficiency $\eta \sim 0.1$.

The spectrum of the high state can be adequately fitted with a single
thermal Comptonization model. The luminosity of the model is $\sim
16.6 \times 10^{39}$ ergs/s and an Amplification factor, $A \sim 2.5$.
The input seed photon luminosity turns out to be $\sim 6.5 \times
10^{40}$ ergs/s. This is significantly higher by a factor of $\sim 4$
of the seed photon luminosity for the low state.  If, like in the low
state, the seed photons are due to an external truncated accretion
disk, the direct emission from such a disk is expected to be $L_{MCD} \sim \frac{L_S}{0.4} \sim 1.6\times10^{41}{\rm~ergs~s^{-1}}$ which would have been
easily detected. The upper-limit to the observed disk emission, however, is only $6.3\times10^{39}{\rm~ergs~s^{-1}}$.  Thus the absence of a soft component in the spectrum argues
against such a geometry. Moreover, if the input seed photon shape is
taken as disk black body emission, the inner disk temperature turns
out to be $0.2$ keV which is similar to that of the low state, despite
the factor of four change in luminosity.  The luminosity of a
truncated accretion disk is $\propto \dot M / R_{tr}$ while the
temperature is $\propto \dot M^{1/4} / R_{tr}^{3/4}$.  Thus a near
constant temperature accompanied by a factor of four increase in
luminosity would imply that the accretion rate $\dot M$ has increased
by a factor of $\sim 8$ while the truncation radius has increased by a
factor of $\sim 2$. The large factor of $\sim 8$  change in the accretion rate
 is not reflected in the change in the total luminosity of
the source, which is only a factor of $\sim 1.5$ higher than the low
state. Thus a truncated accretion disk with an inner hot region does
not seem to be a viable model for the high state of the source.

An alternate geometry for the high state spectrum is that of a corona
covering an optically thick and geometrically thin disk
\citep{1993ApJ...413..507H,1977ApJ...218..247L}.  In this sandwich
model, a fraction of the total gravitational power is dissipated in
the disk, $P_d$ and the rest in the corona $P_c$. The disk photons get
Comptonized by the overlying corona and hence only a thermal
Comptonized spectrum is observed.  A fraction $\xi \la 0.5$ of the
Comptonized photons impinge back on to the underlying disk and are
absorbed. The luminosity of the seed photon is then $L_s = P_d + \xi
L_c$, where $L_c$ is the luminosity of the Comptonized photon.  The
observed total luminosity $L_o = L_c (1-\xi) = P_d+P_c$.  Defining the
Compton amplification factor $A \equiv L_c/L_s$, the dissipated powers
can be related to the total luminosity as,
\begin{equation}
  P_c = \frac{A-1}{(1-\xi)A} L_o
  \label{eqnPc}
\end{equation}
and
\begin{equation}
  P_d  = \frac{1-\xi A}{(1-\xi)A} L_o
  \label{eqnPd}
\end{equation}
The geometry is valid only for $\xi < \xi_{max} = 1/A$. In the absence
of feedback, $\xi = 0$, $P_d = L_o/A = L_d$ and $P_c = L_o - P_d$. In
the other extreme of maximal feedback $\xi = \xi_{max}$, $P_d = 0$ and
$P_c = L_o$.

For the high state, the spectrum can be interpreted in terms of the
corona model, with the seed photon luminosity assumed to be a black
body. The seed photon temperature turns out to $\sim 0.12$ keV and an
amplification factor of $A \sim 2.54$ which imposes an upper limit on
$\xi_{max} \sim 0.4 $.  Thus with a $\xi \la 0.5$ and absence of any
other strong soft component, the spectrum is consistent with the
corona model. The size of the corona region $R_c$ can be estimated
using $\sigma T_s^4 2 \pi R_c^2 = L_s = L_o/(1-\xi)A$. For $L_o \sim
1.7e40$ ergs/sec, $\xi \sim 0.25$ and $A \sim 2.5$, $R_c \sim 3 \times
10^9$ cm.  For a $200 M_\odot$ solar mass black hole this corresponds
to $\sim 50 r_s$. If there is an standard accretion disk beyond this
radius, its temperature would be $< 0.1$ keV and would have a
luminosity $< 2 \times 10^{39}$ ergs/s. Such a emission would not have
been detected. The Eddington ratio for this state turns out be $\sim
0.7$ as compared to $\sim 0.4$ for the low one.
 
\begin{figure}
  \begin{center}
    \includegraphics[width=8.0cm]{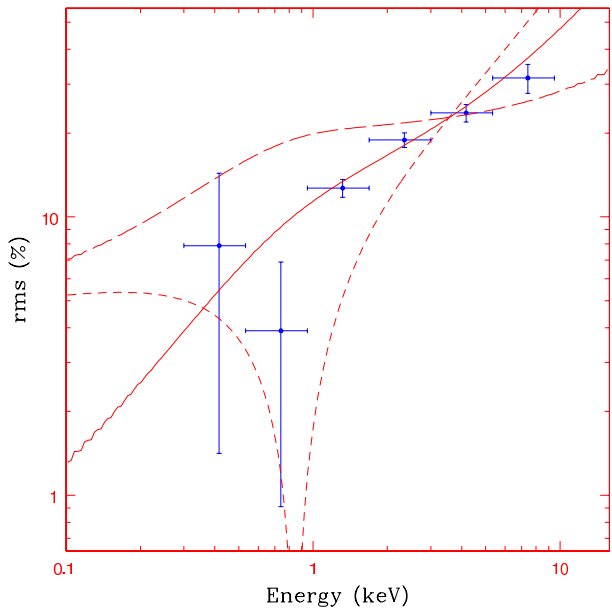}
  \end{center}
  \caption{The observed rms variation with energy for the low state
    data.  The predicted rms variation for $\xi = 0.25$ (solid line),
    $\xi = 0$ (dashed line) and for $\xi = \xi_{max} = 0.33$ (long
    dashed line)}
  \label{figvar}
\end{figure}

The temporal variability of the high state spectrum allows us to
further constrain and verify the corona geometry. The fractional root mean
squared (rms) variability as a function of energy is shown in Figure
(\ref{figvar}).  The simplest interpretation would be that the
observed variability is induced by variations in the coronal power
$P_c$.  From equations (\ref{eqnPc}) and (\ref{eqnPd}), the
amplification can be written as
\begin{equation}
  A (T_e, T_s, \tau)  = \frac{P_d + P_c}{P_d + \xi P_c}
\end{equation}
and the seed photon luminosity as
\begin{equation}
  L_s  = \frac{P_d + \xi P_c}{1-\xi} \propto T_s^4
\end{equation}
A variability in $P_c (t) = P_{c0} (1 + \delta P_c (t))$, will induce
corresponding variation in the amplification, $\delta A$ and the seed
photon temperature $\delta T_s$.  Assuming that the optical depth
$\tau$ of the corona does not vary, the above two equations determine
the time variability of $T_e (t)$ and $T_s(t)$ as a function of
$\delta P_c (t)$. The time dependent spectrum can then be obtained
from $T_e(t)$ and $T_s (t)$.  The standard deviation of the time
dependent spectrum at different energy bands then provides a predicted
rms variation that can be compared with the observed values. For $\xi
= 0.25$, (solid line in Figure \ref{figvar}), the predicted curve is
consistent with the observations. For comparison, curves for $\xi =0$
(dashed line) and $\xi = \xi_{max} = 0.39$ (long dashed line) are also
plotted in the figure.  Note that apart from the overall normalization,
the predicated rms versus energy plot depends only on $\xi$, since
the rest of the parameters are determined by the time-averaged spectral
fitting.  Thus the coronal geometry naturally explains the energy
dependent variability of the source.

As the coronal power increases, the electron temperature should also
increase making the spectrum harder. Thus the model predicts a
correlation between flux and hardness ratio which is indeed seen in
the variability. In the left panel of Figure (\ref{f1}) the hardness
ratio is observed to increase with flux. While the qualitative
behavior is consistent, a quantitative evaluation is unfortunately not
possible. Such a quantitative estimate would require a fairly accurate
knowledge of the response of the instrument to convert the model
photons into counts.  As mentioned earlier, for this observation, the
source extension overlaps with the EPIC-pn chip gap and a large
fraction of events have uncertain pattern.  These events with poor
spectral calibration although excluded from the spectral analysis were
included in the temporal analysis. Excluding them from the temporal
analysis renders the statistics too poor but their inclusion means
that their energies are not well constrained.

The spectrum and rms variability of the source in the high state is
totally consistent with a geometry where a corona covers an underlying
disk in a sandwich model. The photons emitted from the disk are
Comptonized by the corona and only a single thermal component is
seen. A fraction $\xi \sim 0.25$ of the Comptonized photons impinge
back on the disk and are absorbed. The size of the coronal region is
$\sim 50 r_s$ for a $200 M_\odot$ black hole and the Eddington ratio
is $\sim 0.7$.

\section{Summary and Discussion}

We have investigated the spectral and temporal behavior of the bright
ULX NGC~1313 X-1 using two long \xmm{} observations.  On 6 March 2006,
the ULX was in a high flux state with an average count rate of $\sim
1.5$ counts/s, while on in October 2006 the count rate had decreased
to $\sim 0.78$ counts/s. The main results of the analysis are as follows.
\begin{description}
\item The spectrum of the low state shows definite evidence for a
  high energy curvature. It can be fitted with a thermal
  Comptonization model with an optical depth $\tau \sim 15$ and
  temperature $kT_e \sim 2.3 $ keV. An additional soft component,
  which can be modeled as a disk black body emission with a inner disk
  temperature of $kT_{in} \sim 0.23$ keV, is required for the spectral
  fit. The flux from the source is steady with an upper limit on the
  rms variability of $< 3\%$.

\item The spectral shape during the high state is distinctively
  different from that of the low state. The entire spectrum can be
  fitted by a single thermal Comptonization model with an optical
  depth $\tau \sim 13.5$ and temperature $kT_e \sim 1.67 $ keV. There
  is no evidence for any additional soft component.  The source is
  highly variable in this state with an rms variability of $14
  \%$. The associated temporal and spectral difference between the two
  observations indicates that the source underwent a true state
  transition.

\item For a standard radiative efficiency of $\sim 0.1$, the spectrum
  of the low state is completely consistent with a model where a $\sim
  200 M_\odot$ black hole is surrounded by a standard accretion disk
  truncated at a radius of $\sim 17 r_s$ ($r_s \equiv 2GM/c^2$).  This
  outer disk gives rise to the soft component.  There is an inner hot
  region which subtends a solid angle of $\sim 0.8\pi$ to the outer
  disk. Hence photons from the outer disk enter the inner hot region,
  get Comptonized, and are observed as the primary thermal Comptonized
  component. The system has an Eddington ratio of $L_T/L_{Edd}\sim
  0.4$.

\item The spectrum of the high state is not compatible with a
  truncated disk and inner hot region geometry. This is primarily
  inferred by the absence of any strong soft component, despite a four
  fold increase in the input soft photon flux at nearly the same
  temperature. The difference in disk geometry further confirms that a
  true state transition has occurred between the two observations.

\item The spectrum as well as {\it the rms variability as a function
    of energy} of the high state, is completely consistent with a
  model where a geometrically thin disk has a hot corona overlying
  it. Photons from the underlying disk get Comptonized in the corona
  and a single thermal Comptonization model is observed. A fraction
  $\sim 0.25$ of the Comptonized photons impinge back on the disk and
  are absorbed.  The observed variability can be naturally ascribed to
  variations in the coronal power. The size of the coronal region is
  $\sim 50 r_s$ and the Eddington ratio during this state is $\sim
  0.7$.
\end{description}

The low state of NGC 1313 X-1 seems to be analogous to the low hard
state of Galactic X-ray binaries, which can also be broadly explained
in terms of a truncated accretion disk and an hot inner
region. Similarly the coronal geometry for the high state is the same
as the one invoked for the high soft state of Galactic X-ray
binaries. However, this is an important difference between the hot
region/corona of Galactic sources and NGC 1313 X-1.  The hot
region/corona of Galactic sources is optically thin ($\tau \sim 1$)
and has a high temperature, $\sim 50$ keV. The hot region/corona for
NGC 1313 X-1, is optically thick ($\tau \ga 10$) and has a much lower
temperature $\sim 2$ keV. This difference is more than just
quantitative. Optically thin plasmas with electron temperatures $\sim
50$ keV can be in a two-temperature state, with the proton temperature
being much higher (a factor of $\sim 1000$) than that of the electron
gas \citep{1976ApJ...204..187S}. Such plasmas are gas pressure
dominated. Optically thick and low temperature plasmas cannot have
such a two-temperature structure and would be radiation pressure
dominated. Thus, such plasmas would have to be dynamically different
from those observed in Galactic sources.

With an Eddington ratio of $\sim 0.5$, the only difference between NGC
1313 X-1 and Galactic black hole systems seems to be the larger mass
of the black hole. However, the X-ray spectra of bright AGN reveal the
presence of an optically thin and hot plasma similar to those inferred
for Galactic sources. It is not clear why AGN with a $\sim 10^{6-9}
M_\odot$ black hole mass have similar hot region/corona as $\sim 10
M_\odot$ Galactic black holes, but ULX with intermediate masses have
qualitatively different plasmas.  The broadband X-ray spectra of BHBs
in their strongly Comptonized, very high state (VHS) require thermal
Comptonization in an optically thick ($\tau \sim 2-3$) and relatively
cool ($kT_e \sim 10\kev$) plasma in addition to multi-color disk
blackbody emission from a truncated disk and a power-law extending to
high energies \citep{2006MNRAS.371.1216D}. The curvature produced by
the low temperature thermal Comptonization component in BHBs could be
similar to the $\sim 5\kev$ curvature seen in the ULX spectra.
Furthermore, many narrow-line Seyfert 1 galaxies and quasars with high
Eddington ratio show soft X-ray excess emission with $kT\sim 200\ev$
above a hard power-law. The soft excess emission can be modeled with
Thermal Comptonization in an optically thick and cool plasma
\citep[see e.g.,][]{2007ApJ...671.1284D,2008arXiv0807.4847M}. Perhaps the $\sim 5\kev$
curvature observed in ULX, the $\sim 0.2\kev$ soft X-ray excess seen
in many AGN, and the $\sim 10\kev$ thermal Comptonization seen in
strongly Comptonized VHS of BHB are all physically the same continuum
components. There may be an additional hard X-ray ($> 10$ keV)
power-law in ULX which would then be analogous to the X-ray power-law
seen in AGN at $ \gtsim 2$ keV and in VHS of BHBs seen at $\gtsim
10\kev$. However, at this stage such a conjecture would be
speculative.

An alternate to the interpretation presented here is that the source
harbors a $\sim 5 M_\odot$ black hole with super Eddington accretion
\citep{1973A&A....24..337S,2008MNRAS.385L.113K}.  For such high
accretion rates, the luminosity does not scale with $\dot M$, and
instead is given by $L \sim L_{Edd} (1+log(\dot M/\dot
M_E))$. Moreover, the characteristic radius at which the luminosity is
released is $\sim (27/4)(\dot M/\dot M_E)r_s$, which is much larger
than the standard value of $\sim 5 r_s$. The luminosity is also
expected to be further enhanced by some moderate beaming. Such an
interpretation is attractive because (a) it does not require exotic
formation and accretion mechanism needed to explain intermediate mass
black holes, (b) the high accretion rates required are perhaps
expected when there is thermal time scale mass transfer from donor
stars and (c) it explains naturally why the spectra of ULX are
distinct from Galactic sources and AGN which are sub-Eddington
sources. However, the absence of a simple representation of the
expected model spectrum does not allow for detailed spectral analysis.
\cite{2009MNRAS.393L..41K} associate the soft excess seen in some ULX
as emission from such a super Eddington disk. This is reasonable if
the soft excess is the energetically dominant component and not for
cases like the low state of NGC 1313 X-1, where the soft component is
less luminous than the main Comptonization one. Another issue is the
factor of $\sim 2$ luminosity variability between the low and high
state of NGC 1313 X-1. Since luminosity varies as log$(\dot M/\dot
M_E)$ this would would imply a large accretion rate variation which in
turn would imply a large variation in the  characteristic radius
$\propto \dot M$. Such a large variation in radius would result in a
significant change in the characteristic temperature of the disk,
which is not observed. \cite{2009MNRAS.393L..41K} invoke variations in
the moderate beaming factor to explain such observed variations, which
although plausible is rather ad hoc. Finally, if the super Eddington
accretion model is correct, then the self consistent standard disk
modeling which explains both the spectra as well as the variability
versus energy, has to be considered as a coincidence. What is required
is a better theoretical prediction of the super Eddington disk spectra
which may have to include the possibility of a additional dominant
Comptonization component.

In the standard disk framework, there is also a need for better
theoretical understanding of a radiation pressure dominated corona as
well as a hot inner region. The dynamics of such structures needs to be
studied and ascertained that they are viable stable systems. Perhaps
more importantly, the results presented in this work need to be
confirmed by longer observations of the same source as well as other
sources which show similar characteristics. The latter is important
since there may be different kinds of ULX, even among the brightest
ones.

\section*{Acknowledgments}
This work is based on observations obtained with
\xmm{}, an ESA science mission with instruments and contributions
directly funded by ESA Member States and the USA (NASA). This research
has made use of data obtained through the High Energy Astrophysics
Science Archive Research Center Online Service, provided by the
NASA/Goddard Space Flight Center. The long $123\ks$ observation was supported by the NASA grant NNX07AE99G.

\bibliography{ulx}

\label{lastpage}
\end{document}